\documentclass[aps,preprintnumbers, superscriptaddress, amsmath, showpacs]{revtex4}
\usepackage{epsfig}
\usepackage{psfrag}

\usepackage{graphicx}
\usepackage{dcolumn}
\usepackage{bm}

\makeatletter
\renewcommand \thesection {\@arabic\c@section}
\makeatother
\makeatletter 
\@addtoreset{equation}{section}
\makeatother  

\begin{document}


\title{The relativistic correction of the quarkonium melting temperature with a holographic potential}

\author{Yan Wu}
 \email{wuyan@iopp. ccnu. edu. cn}
\affiliation
{Institute of Particle Physics,Key Laboratory of QLP of  MOE, Huazhong Normal
University, Wuhan 430079, China}
\author{ De-fu  Hou}%
 \email{hdf@iopp. ccnu. edu. cn}
 \affiliation{Institute of Particle Physics,Key Laboratory of QLP of  MOE, Huazhong Normal
University, Wuhan 430079, China}
\author{Hai-cang Ren}
\email{ren@mail.rockefeller.edu}   \affiliation{Physics Department, The Rockefeller
University, 1230 York Avenue,  New York,  NY 10021-6399}
\affiliation{Institute of Particle Physics,Key Laboratory of QLP of  MOE, Huazhong Normal
University, Wuhan 430079, China}

\begin{abstract}
The relativistic correction of the AdS/CFT implied heavy quark potential is examined within the framework of the potential
model. For the typical range of the coupling strength appropriate to heavy-ion collisions, we find the correction is
significant in size and lowers the dissociation temperature of quarkonia.
\end{abstract}

\maketitle

\section{Introduction}\label{sec:level1}

The phase structure of quantum chromodynamics (QCD) remains an
active field of researches. At sufficiently high temperature,
hadronic matter will evolve into a quark-gluon plasma which has
been explored experimentally by relativistic heavy ion collisions
(RHIC). The quarkonium dissociation is an important signal of this
transition \cite{Satz}.

The quarkonium are bound states composed of a heavy quark $q$ and
its antiparticle $\bar{q}$. It is found that, the ground states
and the excitation levels of quarkonium are very much smaller than
the normal hadrons, and that they are very tightly
bounded\cite{0512217v2}. Theoretically, there are two approaches
to study the quarkonium: Lattice QCD and potential
models\cite{method}. From Lattice QCD, we can calculate the
spectral function numerically via the quarkonium correlators and
identify the quarkonium states with the resonance
peaks\cite{4,5,6,7,8}. The potential model relies on the small
velocity($v<<1$) of the constituent quarks. By solving a
non-relativistic Schroedinger equation with a temperature
dependent effective potential, we can determine the energy levels
and thereby the threshold temperature when the bound state
dissolves \cite{9,10,11,12,133,14}. The potential model will be
applied in the present work.

AdS/CFT correspondence is a powerful tool to explore the strongly
coupled ${\cal N}=4$ super Yang-Mills plasma. The equation of
state and viscosity ratio, etc. extracted from the AdS/CFT show
remarkable agreement with the Lattice QCD or experimental data
from the quark-gluon plasma (QGP) created via Relativistic Heavy
Ion Collisions (RHIC). It would be interesting to extend the
comparison to a wide range of other quatities, for instance heavy
quark dissociation, which are calculable in both ways to assess
whether the super Yang-Mills serves an important reference model
of the QGP phase of QCD. This is the primary motivation of this
paper.

In a previous work\cite{Hou}, we examined the heavy quarkonium
dissociation within the potential model with the AdS/CFT implied
potential function (holographic potential). We found
that the holographic potential can be approximated by a truncated
Coulomb potential to a great accuracy. With the typical values of
the 't Hooft coupling constant, $\lambda\equiv\sqrt{N_cg_{\rm
YM}^2}$ considered in the literature \cite{Gubser},
\begin{equation}
5.5 < \lambda < 6\pi,
\label{coupling}
\end{equation}
our dissociation temperatures are systematically lower, though not
far from the lattice prediction.
\footnote{The holographic spectral function analysis has been done in \cite{Hoyos}.
The dissociation temperature they obtained is inversely proportional to $\lambda$ and
higher than the Lattice results for $\lambda$ within the domain (\ref{coupling}) .} On the other hand, an estimate of
the velocity of the constituent quarks inside the bound state
indicates that the non-relativistic approximation may be marginal,
especially for $J/\Psi$. This motivates us to examine the
relativistic corrections of the holographic potential with
the aid of a two-body Dirac equation(TBDE).

While the holographic potential alone is sufficient in the nonrelativistic limit, it does not provide all information necessary
for the relativistic corrections even to the order $v^4$
term. Except for the correction brought about by the relativistic kinetic energy, the spin-orbital coupling and the Darwin term depend on how the holographic potential is introduced in the two-body Dirac equation. In addition, gravity dual of spin dependent forces is not available in the literature. Therefore our result remains incomplete at this stage. We would like to comment that the same issues exists for
the relativistic corrections of the heavy quark potential extracted from the lattice QCD simulations.

In the next section, the work reported in \cite{Hou} will be
reviewed and the dissociation temperature beyond the truncated
Coulomb approximation is presented. The corrections to the dissocation temperature are computed in the section 3 through a
Foldy-Wouthuysen (F.W.) transformation of a two-body Dirac
Hamiltonian. The kinetic energy contribution and
the contribution from the Darwin and the spin-orbit coupling are calculated seperately with the latter obtained simply by replacing the perturbative Coulomb potential in the two-body Dirac Hamiltonian with the holographic potential.
The section 4 conclude the paper.

\section{The holographic potential model}

In the conventional potential model of QCD, the
non-relativistic wave function of a heavy quarkonium satisfies the Schroedinger equation
\begin{eqnarray}
[-\frac{1}{2\mu}\nabla^{2}+U(r,T)]\psi=-E(T)\psi,
\label{schroedinger}
\end{eqnarray}
where $E(T)$ is the binding energy and $U(r,T)$ is identified with
the internal energy of a pair a static $q$ and $\bar q$ in QGP and
is related to the free energy $F(r,T)$ via
\begin{eqnarray}
U(r,T)&&=-T^{2}\Big[\frac{\partial}{\partial T}(\frac{F(r,T)}{T})
\Big]_r
\end{eqnarray}
The free energy $F(r,T)$ can be extracted from the expectation of a pair of Wilson loops
operator according to:
\begin{eqnarray}
e^{-\frac{1}{T}F(r,T)}=\frac{tr<W^{\dag}(L_{+})W(L_{-})>}{tr<W^{\dag}(L_{+})><W(L_{-})>}
\label{wilson}
\end{eqnarray}
where $L_{\pm}$ stands for the Wilson loop running in Euclidean time direction at spatial coordinates $(0,0,\pm\frac{1}{2}r)$
and is closed with the periodicity $\beta=\frac{1}{T}$ and
\begin{eqnarray}
W(L_\pm)=Pe^{-i\oint_{L_\pm}dx^{\mu}A_{\mu}(x)}.
\label{loop}
\end{eqnarray}
The spatial coordinates of $L_\pm$ are $(0,0,\pm\frac{1}{2}r)$.
The lattice QCD simulation of the expectation value (\ref{wilson})
can be found if Ref.\cite{10,0746v2}.

In case of super Yang-Mills, the holographic principle places the Wilson lines $L_\pm$ on the boundary ($y\to\infty$) of the 5D AdS-Schwarzschild metric \cite{Maldacena}:
\begin{eqnarray}
ds^{2}=\pi^{2}T^{2}y^{2}(fdt^{2}+d\vec{x}^{2})+\frac{1}{\pi^{2}T^{2}y^{2}f}dy^{2}
\end{eqnarray}
where $f=1-\frac{1}{y^4}$, $d\vec{x}^{2}=dx_1^{2}+dx_2^{2}+dx_3^{2}$ with the ansatz $x_1=x_2=0$ and $x_3$ a function of $y$.

The free energy $F(r,T)$ of the corresponding super Yang-Mills at large $N_c$ and large 't Hooft coupling is proportional to the minimum area of the worldsheet in the AdS
bulk bounded by $L_+$ and $L_-$, and is given parametrically by
\cite{Maldacena,Rey}:
\begin{eqnarray}
\left\{
\begin{aligned}
F(r,T)&=T \min(I,0)\\
r\;\;\;\;&=\dfrac{2q}{\pi T}\int_{y_{c}}^{\infty}\dfrac{dy}{\sqrt{(y^{4}-1)(y^{4}-y_{c}^{4})}}
\label{raduis}
\end{aligned}
\right.
\end{eqnarray}
where
\begin{eqnarray}
I=\sqrt{\lambda}[\int_{y_{c}}^{\infty} dy(
\sqrt{\frac{y^{4}-1}{y^{4}-y_{c}^{4}}}-1)+1-y_{c}]
\label{fenergy}
\end{eqnarray}
and the parameter $y_c\in (1,\infty)$.
Eliminating $y_c$ between (\ref{raduis}) and (\ref{fenergy}), we find that
\begin{eqnarray}
F(r,T)=-\frac{\alpha}{r} \Phi(\rho) \theta(\rho_{0}-\rho)
\end{eqnarray}
where $\alpha\doteq0.2285\sqrt{\lambda}$, $\rho=\pi T r$, $\rho_{0}=0.7541$ and $\Phi(\rho)$ is the screening factor.
The corresponding internal energy is
\begin{eqnarray}
U(r,T)=-\frac{\alpha}{r}[\Phi(\rho)-\rho(\frac{d\Phi}{d\rho})_{y_{c}}-\rho(\frac{d\Phi}{dy_{c}})_{\rho}(\frac{dy_{c}}{d\rho})]\theta(\rho_{0}-\rho)
\label{exact}
\end{eqnarray}
and will be substitute into the Schroedinger equation (\ref{schroedinger}).

The small $\rho$ expansion of $\Phi(\rho)$ is:
\begin{equation}
\Phi(\rho)=1-\frac{\Gamma^{4}(\frac{1}{4})}{4\pi^{3}}\rho+\frac{3\Gamma^{8}(\frac{1}{4})}{640\pi^{6}}\rho^{4}+O(\rho^{8}).
\label{taylor}
\end{equation}
Within the screening radius $\rho_0$, the first two terms of the series (\ref{taylor}) approximate the exact $\Phi$ well as is shown in Fig.1. If we keep only the first two terms, the screening radius $\rho_{0}\simeq0.7359$ and
$U(r,T)$ becomes a truncated Coulomb potential
\begin{eqnarray}
U=-\frac{\alpha}{r}\theta(\rho_{0}-\rho)
\end{eqnarray}
under the approximation.

We define the dissociation temperature $T_d$ as the temperature when the binding
energy falls to zero, i. e. $E(T_d)=0$, and the corresponding radial
Schrodinger equation reads \cite{Hou}:
\begin{eqnarray}
\frac{d^{2}R}{d\rho^{2}}+\frac{2}{\rho}\frac{d
R}{d\rho}-[\frac{l(l+1)}{\rho^{2}}+V]R=0
\end{eqnarray}
where the reduced potential, $V=\frac{mU}{\pi^{2}T^{2}}$ is dimensionless.

The truncated Coulomb potential approximation was employed in \cite{Hou} and the dissociation temperature of the bound state of $l$-th
partial wave and $n$-th radial quantum number is given by
\begin{equation}
T_d=\frac{4\alpha\rho_0m}{\pi x_{nl}^2}
\label{dissociation}
\end{equation}
with $x_{nl}$ the $n$-th nonzero root (ascending order) of the Bessel function $J_{2l}(x)$.
The corresponding radial wave function reads
\begin{equation}
R(r)=\frac{1}{\sqrt{\rho}}J_{2l+1}\left(x_{nl}\sqrt{\frac{\rho}{\rho_0}}\right).
\end{equation}
for $\rho\le\rho_0$ and $R(r)={\rm const.}/r^{l+1}$ for $\rho>\rho_0$.

In this work, we have calculated the dissociation temperature with
the exact holographic potential (\ref{exact}). The comparison with
that obtained from the truncated Coulomb potential in \cite{Hou}
for $J/\Psi$ is shown in Table \uppercase\expandafter{\romannumeral1}, where we
choose $m=1.65$ GeV for the mass of $c$ quarks. From the comparison of these two results we confirmed that the
truncated Coulomb approximation is a good approximation and
we shall stay with the truncated Coulomb approximation for the rest of this paper.
\begin{table}\tabcolsep=17pt 
\begin{minipage}{10cm}{\small{\bf } }
\end{minipage}\vglue7pt
\begin{tabular}{| c | c | c | c | c |}
\hline{} & \multicolumn{2}{|c|} {$T_d$     ($\lambda=5.5$)} & \multicolumn{2}{|c|} {$T_d$     ($\lambda=6\pi$)} \\
\hline{} & {Exact} & {truncated Coulomb} & {Exact} & {truncated Coulomb}\\
\hline{1s} & {142} & {143} & {262} & {265}\\
\hline{2s} & {27} & {27} & {50} & {50}\\
\hline{1p} & {31} & {31} & {57} & {58}\\
\hline
\end{tabular}
\bigskip
\caption{It lists our melting temperature in MeV's for $1s$, $2s$
and $1p$ state using the exact potential and compared with the
Coulomb potential case. We can find that this two results are very
close to each other which confirmed that the Coulomb approximation
is excellent.}
\end{table}

\section{The relativistic correction of the holographic potential}

As it is mentioned in the introduction, the velocity of the heavy
quarks is not low enough so the relativistic correction may be
significant, especially for$J/\Psi$. To explore this correction,
one has to go beyond the Schroedinger equation
(\ref{schroedinger}) and switch to the two body Dirac
equation\cite{Crater,9612445v1,0602066v2,Semay}:
\begin{eqnarray}
i\frac{\partial\Psi}{\partial t}=H\Psi
\end{eqnarray}
In the center-of-mass frame, the Hamiltonian of the two body Dirac
equation is:
\begin{eqnarray}
H=\vec{\alpha_{1}}\cdot\vec{p}+\beta_{1}\cdot
m-\vec{\alpha_{2}}\cdot\vec{p}+\beta_{2}\cdot m+U
\label{TBDE}
\end{eqnarray}
where, $\vec{\alpha}_{1}=\vec{\alpha}\otimes I $, $
\vec{\alpha}_{2}=I\otimes\vec{\alpha} $, $\beta_{1}=\beta\otimes
I$, $\beta_{2}=I\otimes\beta$, $\vec\alpha$, $\beta$ are usual $4\times4$ Dirac matrix, $\vec p=-i\vec\nabla$ and $U$ is the
interaction potential between the two particles. The Hamiltonian
$H$ is a $16\times16$ matrix. A quarkonium state corresponds to a
bound state of $H$ with the eigenvalue $2m-E(T)$, which goes to
$2m$ at the dissociation temperature, i. e. $E(T)=0$. Since we are interested in the
leading order relativistic correction of the dissociation
temperature for the quarkonium, we have to expand the Hamiltonian
to the order $v^{4}$. The sorting of the order in $v$ follows from the
rules that $\frac{\vec{p}^2}{m}\sim U\sim v^2$ and
$\vec\nabla\sim\frac{1}{r}\sim mv$. Also the expectation values of
$\vec\alpha_1$ and $\vec\alpha_2$ are of the order $v$.

In analogous to the one body Foldy-Wouthuysen
transformation\cite{Greiner}, we introduce the unitary
operator
\begin{equation}
{\cal U}=e^{iS_{2}'} e^{iS_{1}'} e^{iS_{2}} e^{iS_{1}}
\end{equation}
where
\begin{eqnarray}
&&S_{1}=-\frac{i}{2m}\beta_{1}\cdot O_{1}\\
&&S_{2}=-\frac{i}{2m}\beta_{2}\cdot (-O_{2})\\
&&S_{1}'=-\frac{i}{2m}\beta_{1}\cdot O_{1}'\ \\
&&S_{2}'=-\frac{i}{2m}\beta_{2}\cdot (-O_{2}')
\end{eqnarray}
with
\begin{eqnarray}
O_{1}=\vec{\alpha}_{1}\cdot\vec{p},\ \ \ \
O_{2}=\vec{\alpha}_{2}\cdot\vec{p}.\nonumber
\end{eqnarray}
The transformed Hamiltonian reads
\begin{eqnarray}
H_{\rm FW}&=&{\cal U}H{\cal U}^\dagger\nonumber \\
&=&(\beta_{1}+\beta_{2})(m+\frac{\vec{p}^{2}}{2m}-\frac{\vec{p}^{4}}{8m^{3}})+U
+\frac{1}{4m^{2}}\nabla^2{U}+\frac{1}{4m^{2}r}\frac{dU}{dr}(\vec{\sigma}_{1}+\vec{\sigma}_{2})\cdot\vec{L}
\label{Hfw}
\end{eqnarray}
where, higher order terms in $v$ have been dropped. The details of
the transformation are deferred to the appendix. The
non-relativistic wave function,
$\Psi=\Psi_{s_1s_2}(\vec{r}_1-\vec{r}_2)$ with subscript
$s_1, s_2$ labelling the spin components of the two
quarks, corresponding to the sector with $\beta_1=\beta_2=1$
(This wave function can be expanded in the series of products of
the orbital wave functions of the preceding section and the spin wave functions). We may stay within this sector for
the 1st order perturbation of the $v^4$ term of (\ref{Hfw}) with
the effective Hamiltonian $H_{\rm eff.}=H_0+H_1$, where
\begin{eqnarray}
H_{0}=2m+\frac{\vec{p}^{2}}{m}+U \nonumber
\end{eqnarray}
corresponds to the non-relativistic part,\\
and
\begin{eqnarray}
H_1 &=& -\frac{\vec{p}^{4}}{4m^{3}}+\frac{1}{4m^{2}}\nabla^2{U}+\frac{1}{4m^{2}r}\frac{dU}{dr}(\vec{\sigma}_{1}+\vec{\sigma}_{2})\cdot \vec{L}\nonumber\\
&=& -\frac{\vec{p}^{4}}{4m^{3}}+\frac{1}{4m^{2}}\nabla^2{U}
+\frac{1}{4m^{2}r}\frac{dU}{dr}(J^2-L^2-S^2), \label{h1}
\end{eqnarray}
is a relativistic correction. We have introduced the total spin $\vec
S=\frac{1}{2}(\vec\sigma_1+\vec\sigma_2)$ and the total angular
momentum $\vec J=\vec L+\vec S$ in the last step. The contribution
of $H_1$ is somewhat like the that to the fine structure of the
hydrogen atom, and we can see that the first term is the first
order of the kinetic energy correction; the second term is Drawin
term; and the third term is the spin orbit coupling which
can be decomposed into the spin singlet and the spin triplet channels.

Perturbatively, we may write $E(T)=E_0(T)+\delta E(T)$ and
$T=T_0+\delta T$, where $E_0(T)$ is the non-relativistic binding
energy in (\ref{schroedinger}), and $T_0$ are the dissociation temperature given by
(\ref{dissociation}), and $\delta E(T)$ and $\delta T$ are the
$v^4$ corrections. We have $E_0(T_0)=0$. Expanding the
dissociation condition $E(T)=0$ to the order $v^4$, we obtain the formula for $\delta T$, i.e.
\begin{equation}
\delta T=\delta_1 T+\delta_2 T=-\frac{\delta_1 E(T_0)+\delta_2 E(T_0)}{(\frac{\partial E_0}{\partial T})_{T_0}}.
\end{equation}
where,
\begin{eqnarray}
\delta_1 E(T_0)&=&-<\frac{\vec{p}^{4}}{4m^{3}}>\nonumber\\
\delta_2 E(T_0)&=&\Big[<\frac{1}{4m^{2}}\nabla^2{U}>+<\frac{1}{4m^{2}r}\frac{dU}{dr}(J^2-L^2-S^2)>]
\end{eqnarray}
and
\begin{equation}
\left(\frac{\partial E_0}{\partial
T}\right)_{T=T_0}=<\frac{\partial H_0}{\partial
T}>=<\frac{\partial U}{\partial T}>,
\end{equation}
The average
\begin{equation}
<O>\equiv\frac{\int d^3\vec r\psi^*(\vec r)O(\vec r)\psi(\vec r)}{\int d^3\vec r\psi^*(\vec r)\psi(\vec r)}.
\end{equation}
with $\psi(\vec r)$ the non-relativistic wave function.
The reason for our separating the contribution from $p^4$, $\delta_1 T$ and that from the Darwin and spin-orbital terms, $\delta_2 T$ is the uncertainty in the representation of the holographic potential in (\ref{TBDE}), which does not impact on the $p^4$ correction. We will come to this point in the next section.  In the limit of zero binding energy, we find
$<\frac{\vec{p}^{4}}{4m^{3}}>=<\frac{1}{4m}U^2>$. For the
truncated Coulomb potential,
\begin{equation}
\nabla^{2}U=4\pi\alpha\pi^{3}T^{3}\delta^{3}(\vec{\rho})\theta(\rho_{0}-\rho)+\frac{\alpha \pi^{3}T^{3}}{\rho}\delta'(\rho-\rho_{0}).
\end{equation}

In terms of the radial wave function $R_l(r)$ of $\psi(\vec r)$,
\begin{eqnarray}
<-\frac{\vec{p}^{4}}{4m^{3}}>&&=-\frac{\alpha^{2}}{4m\pi T}\int_{0}^{\rho_{0}}R_l(\rho)^{2}d\rho\nonumber \\
<\frac{1}{4m^{2}}\nabla^2{U}>&&=\frac{\alpha}{4m^{2}}\{|R_l(0)|^{2}-2R_l(\rho_{0})R_l'(\rho_{0})\rho_{0}-R_l^{2}(\rho_{0})\}\nonumber \\
<\frac{1}{4m^2r}\frac{d
U}{dr}>&&=\frac{\alpha}{4m^2}\int_0^{\rho_0}\frac{d\rho}{\rho}R^2(\rho)
+\frac{\alpha}{4m^2}R^2(\rho_0)\nonumber \\
<\frac{\partial U}{\partial T}>&&=\frac{\alpha\pi}{\pi^{3}T^{3}}R_l^{2}(\rho_{0})\rho_{0}^{2}
\end{eqnarray}
For the $ns$ state, we find that
\begin{eqnarray}
&&\delta_1 T=\frac{\pi\alpha T_0^2}{4m\rho_0}\Big[\frac{1}{J_1^2(x_{n0})}-\frac{J_0^2(x_{n0})}{J_1^2(x_{n0})}-1\Big]\nonumber\\
&&\delta_1 T+\delta_2 T=-\frac{\pi\alpha T_0^2}{4m\rho_0}\Big[\frac{J_0^2(x_{n0})}{J_1^2(x_{n0})}+1+\frac{2}{x_{n0}^2}\Big].
\end{eqnarray}
For the $np$ state, we can also get analytical expressions, which are more lengthy.\\

The numerical values of the corrected temperature $T_0+\delta_1 T$ and $T_0+\delta_1 T+\delta_2 T$ in MeV's for 1$s$, 2$s$ and
1$p$ states are listed in the Table
\uppercase\expandafter{\romannumeral2} below.
\begin{table}
\tabcolsep=17pt  
\small
\renewcommand\arraystretch{2.5}  
\begin{minipage}{10cm}{
\small{\bf } }
\end{minipage}
\vglue7pt
\begin{tabular}{| c | c | c | c | c |}
\hline
 \multicolumn{1}{|c|} {}&\multicolumn{2}{|c|} {$c\bar{c}$}& \multicolumn{2}{|c|} {$b\bar{b}$}\\ \cline{2-5}

 \multicolumn{1}{|c|} {}& {$\lambda=5.5$} &  {$\lambda=6\pi$}& {$\lambda=5.5$} &  {$\lambda=6\pi$} \\
\hline
 {1$s$}& {162.54} & {387.54}&{478.76}&{1139.11}\\
\hline
 {2$s$}& {29.15} & {62.75}&{85.67}&{184.44}\\
\hline
 {1$p$}& {32.04} & {62.14}&{94.18}&{182.66}\\
\hline
\end{tabular}
\end{table}

\begin{table}
\tabcolsep=17pt  
\small
\renewcommand\arraystretch{2.5}  
\begin{minipage}{10cm}{
\small{\bf } }
\end{minipage}
\vglue7pt
\begin{tabular}{| c | c | c | c | c |}
\hline
 \multicolumn{1}{|c|} {}&\multicolumn{2}{|c|} {$c\bar{c}$}& \multicolumn{2}{|c|} {$b\bar{b}$}\\ \cline{2-5}

 \multicolumn{1}{|c|} {}& {$\lambda=5.5$} &  {$\lambda=6\pi$}& {$\lambda=5.5$} &  {$\lambda=6\pi$} \\
\hline
 {1$s^{1}_{0}$}& {130.79} & {188.65}&{385.63}&{555.58}\\
 \hline
{1$s^{3}_{1}$}& {130.79} & {188.65}&{385.63}&{555.58}\\
\hline
 {2$s^{1}_{0}$}& {26.71} & {48.16}&{79.15}&{142.59}\\
\hline
 {2$s^{3}_{1}$}& {26.71} & {48.16}&{79.15}&{142.59}\\
\hline
 {1$p^{1}_{1}$}& {31.53} & {61.33}&{93.54}&{180.79}\\
\hline
 {1$p^{3}_{0}$}& {32.65} & {68.48}&{96.85}&{201.80}\\
\hline
 {1$p^{3}_{1}$}& {32.09} & {64.90}&{95.20}&{191.30}\\
\hline
 {1$p^{3}_{2}$}& {30.96} & {57.76}&{91.89}&{170.29}\\
\hline
\end{tabular}
\bigskip
\caption{This lists the final results of the corrected temperature
in MeV's, where the upper one corresponds to the results of
$T_0+\delta_1 T$ and the lower one corresponds to $T_0+\delta_1
T+\delta_2 T$. For the lower one, we wrote the states as
$nL^{2S+1}_{J}$ in the first column, where $n$ is the main quantum
number, $L$ is the orbit angular momentum quantum number, $S$ is
the spin quantum number and $J$ is the total angular momentum
quantum number. Since there are no spin-orbit coupling term for
$ns$ states, the spin singlet and spin triplet are degenerate.
However, for $p$ state, adding the coupling term, there will have
energy level splitting, so we can see the different results for
different total angular momentum $J$.}
\end{table}

\section{Discussions}

In summary, we have explored the leading relativistic correction
to the dissociation temperature of heavy quarkonium state through
a F.W.-like transformation of the two body Dirac Hamiltonian with
the AdS/CFT implied potential. Among the contributions we
considered, the $p^4$ correction of the kinetic energy, being
negative, enhances the binding but the Darwin term does the
opposite and dominates. Consequently, the dissociation temperature
of $s$-state is lowered, leaving the corrected values further
below the lattice result. This disagreement can be attributed to
the short screening length $r_0=\frac{\rho_0}{\pi T}$, about 0.25fm at $T=200$ MeV, of the
AdS/CFT potential and the sharp cutoff nature of the screening. In
case of $J/\psi$, the magnitude of the correction ranges from
$8\%$ for $\lambda=5.5$ to $30\%$ for $\lambda=6\pi$, indicating
significant relativistic towards the high end of the domain
(\ref{dissociation}).

The potential model, though physically more transparent than
spectral function approach, does not provide complete $v^4$
corrections with the holographic potential extracted from the
Wilson loop alone. The same deficiency applies the relativistic correction based on the lattice heavy quark potential alone. As the Wilson loop for a nonAbelian theory involves multi-gluon excahnges, its form in the two-body Dirac Hamiltonian (\ref{TBDE}) may not be adequate unless the single gluon exchange serves a reasonable approximation. Among the four
sectors $\beta_1=\pm 1$ and $\beta_2=\pm 1$, only two of them
$\beta_1=\beta_2=\pm 1$ correspond to $q\bar q$ interacting via
the holographic potential. The other two sectors with $\beta_1$
and $\beta_2$ correspond to $qq$ or $\bar{q}\bar{q}$ and the
interacting potential is unknown. A more general form of the
interaction in (\ref{TBDE}) without violating the charge conjugation symmetry is to replace $U$ by
\begin{equation}
(\Lambda_{++}+\Lambda_{--})U+(\Lambda_{+-}+\Lambda_{-+})U'=U_++\beta_1\beta_2U_-
\end{equation}
where the projection operator $\Lambda_{\pm 1,\pm 1}\equiv\frac{1\pm\beta_1}{2}
\frac{1\pm\beta_2}{2}$, $U'$ the potential between $qq$ or
$\bar q\bar q$ and $U_\pm=\frac{U\pm U'}{2}$. $U'=U$ in the previous section.
Repeating the steps of FW transformation in the appendix,
we find that the perturbing Hamiltonian (\ref{h1}) is replaced by
\begin{equation}
H_1=-\frac{\vec{p}^{4}}{4m^{3}}+\frac{1}{4m^{2}}\nabla^2{U_+}
+\frac{1}{4m^{2}r}\frac{dU_+}{dr}(\vec{\sigma}_{1}+\vec{\sigma}_{2})\cdot
\vec{L}
+\frac{1}{8m^2}(\lbrace\vec\sigma_1\cdot\vec\nabla,\lbrace\vec\sigma_1\cdot\vec\nabla,U_-\rbrace\rbrace
+\lbrace\vec\sigma_2\cdot\vec\nabla,\lbrace\vec\sigma_2\cdot\vec\nabla,U_-\rbrace\rbrace
\end{equation}
with $\lbrace...\rbrace$ an anticommutator. This will modify the potential part of (\ref{h1}). Only the correction from the
kinetic energy, $p^4$ term of (\ref{h1}) is robust, which raise
the dissociation temperature.

In addition to the holographic potential considered in this work, there should be spin dependent ones
that splits degeracies between spin singlets and spin triplets (e.g. between $\eta_c$ and $J/\psi$). The
gravity dual of the latter are unknown in the literature.
A first principle derivation of the spin-dependent forces associate them to
the expectation value of Wilson loops with operator insertions, $<{\rm tr}{\cal
W}_{\mu\nu}(L_+)^\dagger {\cal W}_{\rho\lambda}(L_-)>$, where
\begin{equation}
{\cal W}_{\mu\nu}(L)=PF_{\mu\nu}(x)e^{-i\int_Ldx^\mu A_\mu}
\end{equation}
with $F_{\mu\nu}$ the Yang-Mills field strength and $x$ a point along $L$, ${\cal W}_{\mu\nu}(L)$ is obtained
from the Eqs.(\ref{loop}) by small distortion of $L$ at $x$. Within the AdS/CFT framework, it corresponds to
the perturbation of the Nambu-Goto action of the world shee underlying the holographic potential under a
small distortion of its boundary. It is a challenging boundary
value problem and we hope to report our progress along this line in future.

Finally, we would like to comment on a phenomenological formulation of the two-body Dirac equation
\cite{wong}, which has been applied recently to the same problem addressed in this work \cite{zhuang}. It amounts to
divide the heavy quark potential of Cornell type into a linearly confining term and a single gluon Coulomb term and
to generate all spin-dependent forces by the latter. While it is legitimate in vacuum in the weak coupling because of the Lorentz
invariance, a direct application to a medium beyond weak coupling remains to be justified, given different screening properties of the electric and magnetic gluons.

\section{Acknowlegdment}\label{sec:level1}
We thank Prof. Pengfei Zhuang for bringing our attention to their recent paper \cite{zhuang} prior to uploading to the web.
This work is supported in part by NSFC under grant  Nos. 10975060, 11135011, 11221504

\appendix
\section{}

In this appendix, we shall fill in the details of the F.W.
transformation for the two body Dirac equation which we have done
in section 4.

Let us recall the Foldy-Wouthuysen transformation of the one body Dirac Hamiltonian,
$H=\vec\alpha\cdot\vec p+\beta m+V$. For a 4-component spinor with velocity $v<<1$ and positive
energy, one can work in a representation where the upper two component corresponds to the
non-relativistic limit, referred to as the large components, while the lower two components
are suppressed by a power of $v$, referred to as the small components. An operator is even(odd)
if it is diagonal(off-diagonal) with respect to large and small components. For example, $\beta$
is even and $\vec\alpha$ is odd. The Foldy-Wouthuysen transformation amounts to successive
unitary transformations that push the odd operators to higher orders in $v$.

The Foldy-Wouthuysen transformation can be easily generalized to the two body case, the
Hilbert space of which is spanned by the direct products of two one body spinors.
To the leading order relativistic correction, we stop at the $v^4$ terms ignoring all the
higher order term.
The Hamiltonian of the two body Dirac equation is:
\begin{eqnarray}
H=\vec{\alpha_{1}}\vec{p}+\beta_{1}m-\vec{\alpha_{2}}\vec{p}+\beta_{2}m+V
\end{eqnarray}
let:
\begin{eqnarray}
\vec{\alpha_{1}}\vec{p}=O_{1}\nonumber\\
\vec{\alpha_{2}}\vec{p}=O_{2}\nonumber
\end{eqnarray}
so,
\begin{eqnarray}
H&&=\underbrace{\beta_{1}m+O_{1}}+\underbrace{\beta_{2}m-O_{2}}+V\nonumber\\
&&=H_{1}+H_{2}+V\nonumber\\
\end{eqnarray}
where $H_{1}$ corresponds to the first under-brace, and $H_{2}$
corresponds to the second under-brace.
And it is easy to proof the commutation relation that:$[O_{1}, O_{2}]=[\beta_{1}, \beta_{2}]=[O_{1}, \beta_{2}]=[O_{2}, \beta_{1}]=0$, and $O_{1}\sim O_{2}\sim v$.\\
We do the first transformation:\\
we select:
\begin{eqnarray}
S_{1}&&=-\frac{i}{2m}\beta_{1}O_{1}\nonumber\\
S_{2}&&=-\frac{i}{2m}\beta_{2}(-O_{2})\nonumber
\end{eqnarray}
so, the transformed Hamiltonian turns into:
$H'=\underbrace{e^{iS_{2}}\overbrace{e^{iS_{1}}He^{-iS_{1}}}e^{-iS_{2}}}$.
We calculate the mid over-brace first.
\begin{eqnarray}
e^{iS_{1}}He^{-iS_{1}}&&=\underbrace{e^{iS_{1}}H_{1}e^{-iS_{1}}}+\underbrace{e^{iS_{1}}H_{2}e^{-iS_{1}}}+\underbrace{e^{iS_{1}}Ve^{-iS_{1}}}\nonumber\\
&&=\overline{H_{1}}+\overline{H_{2}}+V'\nonumber\\
\end{eqnarray}
also, $\overline{H_{1}}$ corresponds to the first under-brace,
$\overline{H_{2}}$ corresponds to the second, and $V'$ corresponds
to the third.
\begin{eqnarray}
\overline{H_{1}}&&=H_{1}+[iS_{1}, H_{1}]+\frac{1}{2!}[iS_{1}, [iS_{1}, H_{1}]]+\frac{1}{3!}[iS_{1}, [iS_{1}, [iS_{1}, H_{1}]]]+\frac{1}{4!}[iS_{1}, [iS_{1}, [iS_{1}, [iS_{1}, H_{1}]]]]+\cdots\\
&&[iS_{1}, H_{1}]=-O_{1}+\frac{1}{m}\beta_{1}O_{1}^{2}\nonumber\\
&&\frac{1}{2!}[iS_{1}, [iS_{1}, H_{1}]]=-\frac{1}{2m}\beta_{1}O_{1}^{2}-\frac{1}{2m^{2}}O_{1}^{3}\nonumber\\
&&\frac{1}{3!}[iS_{1}, [iS_{1}, [iS_{1}, H_{1}]]]=\frac{1}{6m^{2}}O_{1}^{3}-\frac{1}{6m^{3}}\beta_{1}O_{1}^{4}\nonumber\\
&&\frac{1}{4!}[iS_{1}, [iS_{1}, [iS_{1}, [iS_{1}, H_{1}]]]]\sim\frac{1}{24m^{3}}\beta_{1}O_{1}^{4}\nonumber\\
\overline{H_{1}}&&=\beta_{1}m+\frac{1}{2m}\beta_{1}O_{1}^{2}-\frac{1}{8m^{3}}\beta_{1}O_{1}^{4}-\frac{1}{3m^{2}}O_{1}^{3}+O(v^{5})
\end{eqnarray}
where $v$ is the velocity of the particles.
\begin{eqnarray}
\overline{H_{2}}&&=H_{2}+[iS_{1}, H_{2}]+\frac{1}{2!}[iS_{1}, [iS_{1}, H_{2}]]+\frac{1}{3!}[iS_{1}, [iS_{1}, [iS_{1}, H_{2}]]]+\frac{1}{4!}[iS_{1}, [iS_{1}, [iS_{1}, [iS_{1}, H_{2}]]]]+\cdots\\
&&[iS_{1}, H_{2}]=\frac{1}{2!}[iS_{1}, [iS_{1}, H_{2}]]=\frac{1}{3!}[iS_{1}, [iS_{1}, [iS_{1}, H_{1}]]]=\cdots=0\nonumber\\
\overline{H_{2}}&&=H_{2}=\beta_{2}m-O_{2}\nonumber\\
V'&&=V+\frac{1}{2m}\beta_1[O_1,V]-\frac{1}{8m^2}[O_1,[O_1,V]]
\end{eqnarray}
so,
\begin{eqnarray}
e^{iS_{1}}He^{-iS_{1}}&&=\overline{H_{1}}+\overline{H_{2}}+V'\nonumber\\
&&=\beta_{1}m+\beta_{2}m-O_{2}+V+\frac{1}{2m}\beta_{1}O_{1}^{2}-\frac{1}{8m^{2}}[O_{1}, [O_{1}, V]]-\frac{1}{8m^{3}}\beta_{1}O_{1}^{4}+\frac{1}{2m}\beta_{1}[O_{1}, V]-\frac{1}{3m^{2}}O_{1}^{3}+O(v^{5})\nonumber\\
&&=\overline{H}
\end{eqnarray}
that we mark $e^{iS_{1}}He^{-iS_{1}}$ as $\overline{H}$ for convenience.
So,
\begin{eqnarray}
H'&&=e^{iS_{2}}\overline{H}e^{-iS_{2}}\nonumber\\
&&=\overline{H}+[iS_{2}, \overline{H}]+\frac{1}{2!}[iS_{2}, [iS_{2}, \overline{H}]]+\frac{1}{3!}[iS_{2}, [iS_{2}, [iS_{2}, \overline{H}]]]+\frac{1}{4!}[iS_{2}, [iS_{2}, [iS_{2}, [iS_{2}, \overline{H}]]]]+\cdots\\
&&[iS_{2}, \overline{H}]=O_{2}+\frac{1}{m}\beta_{2}O_{2}^{2}-\frac{1}{2m}\beta_{2}[O_{2}, V]-\frac{1}{4m}\beta_{1}\beta_{2}[O_{2}, [O_{1}, V]]\nonumber\\
&&\frac{1}{2!}[iS_{2}, [iS_{2},\overline{H}]]=-\frac{1}{2m}\beta_{2}O_{2}^{2}+\frac{1}{2m^{2}}O_{2}^{3}-\frac{1}{8m^{2}}[O_{2}, [O_{2}, V]]\nonumber\\
&&\frac{1}{3!}[iS_{2}, [iS_{2}, [iS_{2}, \overline{H}]]]=-\frac{1}{6m^{2}}O_{2}^{3}-\frac{1}{6m^{3}}\beta_{2}O_{2}^{4}\nonumber\\
&&\frac{1}{4!}[iS_{2}, [iS_{2}, [iS_{2}, [iS_{2}, \overline{H}]]]]\sim\frac{1}{24m^{3}}\beta_{2}O_{2}^{4}\nonumber\\
H'=&&\beta_{1}m+\beta_{2}m+V\nonumber\\
&&+\frac{1}{2m}\beta_{1}O_{1}^{2}-\frac{1}{8m^{2}}[O_{1}, [O_{1}, V]]-\frac{1}{8m^{3}}\beta_{1}O_{1}^{4}+\frac{1}{2m}\beta_{1}[O_{1}, V]-\frac{1}{3m^{2}}O_{1}^{3}\nonumber\\
&&+\frac{1}{2m}\beta_{2}O_{2}^{2}-\frac{1}{8m^{2}}[O_{2}, [O_{2}, V]]-\frac{1}{8m^{3}}\beta_{2}O_{2}^{4}-\frac{1}{2m}\beta_{2}[O_{2}, V]+\frac{1}{3m^{2}}O_{2}^{3}\nonumber\\
&&-\frac{1}{4m^{2}}\beta_{1}\beta_{2}[O_{2}, [O_{1}, V]]+O(v^{5})
\end{eqnarray}
since,
\begin{eqnarray}
H'&&=(\beta_{1}m+V+\frac{1}{2m}\beta_{1}O_{1}^{2}-\frac{1}{8m^{2}}[O_{1}, [O_{1}, V]]-\frac{1}{8m^{3}}\beta_{1}O_{1}^{4}+\frac{1}{2m}\beta_{1}[O_{1}, V]-\frac{1}{3m^{2}}O_{1}^{3})\Longrightarrow mark:H_{1}'\nonumber\\
&&+(\beta_{2}m+\frac{1}{2m}\beta_{2}O_{2}^{2}-\frac{1}{8m^{2}}[O_{2}, [O_{2}, V]]-\frac{1}{8m^{3}}\beta_{2}O_{2}^{4}-\frac{1}{2m}\beta_{2}[O_{2}, V]+\frac{1}{3m^{2}}O_{2}^{3})\Longrightarrow mark:H_{2}'\nonumber\\
&&-\frac{1}{4m^{2}}\beta_{1}\beta_{2}[O_{2}, [O_{1}, V]]\\
&&H_{1}'=\beta_{1}m+\underbrace{V+\frac{1}{2m}\beta_{1}O_{1}^{2}-\frac{1}{8m^{2}}[O_{1}, [O_{1}, V]]-\frac{1}{8m^{3}}\beta_{1}O_{1}^{4}}+\underbrace{\frac{1}{2m}\beta_{1}[O_{1}, V]-\frac{1}{3m^{2}}O_{1}^{3}}\nonumber\\
&&=\beta_{1}m+V_{1}'+O_{1}'\\
&&H_{2}'=\beta_{2}m+\underbrace{\frac{1}{2m}\beta_{2}O_{2}^{2}-\frac{1}{8m^{2}}[O_{2}, [O_{2}, V]]-\frac{1}{8m^{3}}\beta_{2}O_{2}^{4}}-\underbrace{(\frac{1}{2m}\beta_{2}[O_{2}, V]-\frac{1}{3m^{2}}O_{2}^{3})}\nonumber\\
&&=\beta_{2}m+V_{2}'+O_{2}'\\
H'&&=H_{1}'+H_{2}'-\frac{1}{4m^{2}}\beta_{1}\beta_{2}[O_{2},
[O_{1}, V]]
\end{eqnarray}
the first under-braces upside in the expression of $H_{1}'(H_{2}')$ correspond to $V_{1}'(V_{2}')$, and the second correspond to $O_{1}'(O_{2}')$ for convenience.\\
Then, we do the second transformation:\\
We select:
\begin{eqnarray}
S_{1}'&&=-\frac{i}{2m}\beta_{1}O_{1}'\nonumber\\
S_{2}'&&=-\frac{i}{2m}\beta_{2}(-O_{2}')\nonumber
\end{eqnarray}
and $O'_{1}\sim O'_{2}\sim v^3$\\
so,
\begin{eqnarray}
H''&&=e^{iS_{2}'}e^{iS_{1}'}H'e^{-iS_{1}'}e^{-iS_{2}'}\nonumber\\
&&=e^{iS_{2}'}\underbrace{e^{iS_{1}'}H_{1}'e^{-iS_{1}'}}e^{-iS_{2}'}+e^{iS_{2}'}\underbrace{e^{iS_{1}'}H_{2}'e^{-iS_{1}'}}e^{-iS_{2}'}-\frac{1}{4m^{2}}e^{iS_{2}'}e^{iS_{1}'}\beta_{1}\beta_{2}[O_{2}, [O_{1}, V]]e^{-iS_{1}'}e^{-iS_{2}'}\nonumber\\
&&=e^{iS_{2}'}\overline{H_{1}'}e^{-iS_{2}'}+e^{iS_{2}'}\overline{H_{2}'}e^{-iS_{2}'}-\frac{1}{4m^{2}}e^{iS_{2}'}e^{iS_{1}'}\beta_{1}\beta_{2}[O_{2}, [O_{1}, V]]e^{-iS_{1}'}e^{-iS_{2}'}
\end{eqnarray}
where, the first under-brace upside correspond to $\overline{H_{1}'}$, and the second correspond to $\overline{H_{2}'}$,
and we can directly apply the results of the first transformation\\
\begin{eqnarray}
\overline{H_{1}'}&&=H_{1}'+[iS_{1}', H_{1}']+O(v^5)\nonumber\\
&&=\beta_{1}m+V_{1}'+\frac{1}{2m}\beta_{1}[O_{1}', V_{1}']
\end{eqnarray}
\begin{eqnarray}
e^{iS_{2}'}\overline{H_{1}'}e^{-iS_{2}'}&&=\overline{H_{1}'}+[iS_{2}', \overline{H_{1}'}]+O(v^5)\nonumber\\
&&=\beta_{1}m+V_{1}'+\frac{1}{2m}\beta_{1}[O_{1}',
V_{1}']-\frac{1}{2m}\beta_{2}[O_{2}', V_{1}']
\end{eqnarray}
\begin{eqnarray}
\overline{H_{2}'}&&=H_{2}'+[iS_{1}', H_{2}']+O(v^5)\nonumber\\
&&=\beta_{2}m+V_{2}'-O_{2}'+\frac{\beta_{1}}{2m}[O_{1}',V_{2}']-\frac{\beta_{1}}{2m}[O_{1}',O_{2}']
\end{eqnarray}
\begin{eqnarray}
e^{iS_{2}'}\overline{H_{2}'}e^{-iS_{2}'}&&=\overline{H_{2}'}+[iS_{2}', \overline{H_{2}'}]+O(v^5)\nonumber\\
&&=\beta_{2}m+V_{2}'-\frac{1}{2m}\beta_{2}[O_{2}',
V_{2}']+\frac{1}{2m}\beta_{1}[O_{1}',
V_{2}']-\frac{1}{2m}\beta_{1}[O_{1}', O_{2}']
\end{eqnarray}
Then, we see the last term in $H''$: $-\frac{1}{4m^{2}}\underbrace{e^{iS_{2}'}\overbrace{e^{iS_{1}'}\beta_{1}\beta_{2}[O_{2}, [O_{1}, V]]e^{-iS_{1}'}}e^{-iS_{2}'}}$\\
we do the calculation of the over-brace first.
\begin{eqnarray}
e^{iS_{1}'}\beta_{1}\beta_{2}[O_{2}, [O_{1}, V]]e^{-iS_{1}'}&&=\beta_{1}\beta_{2}[O_{2}, [O_{1}, V]]+[iS_{1}',\beta_{1}\beta_{2}[O_{2}, [O_{1}, V]]]+O(v^5)\nonumber\\
&&=(\beta_{1}\beta_{2}[O_{2}, [O_{1},
V]]+\frac{1}{4m^{2}}[\beta_{1}[O_{1},V],[O_{2},[O_{1},V]]])\Longrightarrow
mark:A
\end{eqnarray}
\begin{eqnarray}
e^{iS_{2}'}Ae^{-iS_{2}'}&&=A+O(v^5)\\
&&=-\frac{1}{4m^{2}}\beta_{1}\beta_{2}[O_{2}, [O_{1}, V]]
\end{eqnarray}
considering $O_{1},O_{2}\sim v, O_{1}',O_{2}'\sim v^{3}$
\begin{eqnarray}
H''=&&\beta_{1}m+V_{1}'+\beta_{2}m+V_{2}'-\frac{1}{4m^{2}}\beta_{1}\beta_{2}[O_{2}, [O_{1}, V]]\nonumber\\
=&&\beta_{1}m+\beta_{2}m+V+\frac{1}{2m}\beta_{1}O_{1}^{2}+\frac{1}{2m}\beta_{2}O_{2}^{2}-\frac{1}{8m^{2}}[O_{1},[O_{1},V]]-\frac{1}{8m^{2}}[O_{2},[O_{2},V]]\nonumber\\
&&-\frac{1}{8m^{3}}\beta_{1}O_{1}^{4}-\frac{1}{8m^{3}}\beta_{2}O_{2}^{4}-\frac{1}{4m^{2}}\beta_{1}\beta_{2}[O_{2}, [O_{1}, V]]\\
since,&&O_{1}^{2}=O_{2}^{2}=\vec{p}^{2},O_{1}^{4}=O_{2}^{4}=\vec{p}^{4}\nonumber\\
&&[O_{1},[O_{1},V]]=-\nabla^{2}V-\frac{2}{r}\frac{\partial V}{\partial r}\vec{\Sigma_{1}}\vec{L}\nonumber\\
&&[O_{2},[O_{2},V]]=-\nabla^{2}V-\frac{2}{r}\frac{\partial V}{\partial r}\vec{\Sigma_{2}}\vec{L}\nonumber\\
&&[O_{2},[O_{1},V]]=-\alpha_{1i}\alpha_{2j}\nabla_{i}\nabla_{j}V\nonumber\\
H''=&&\beta_{1}(m+\frac{\vec{p}^{2}}{2m}-\frac{\vec{p}^{4}}{8m^{3}})+\beta_{2}(m+\frac{\vec{p}^{2}}{2m}-\frac{\vec{p}^{4}}{8m^{3}})+V\nonumber\\
&&+\frac{1}{4m^2}\nabla^2V
+\frac{1}{4m^2}\frac{1}{r}\frac{\partial V}{\partial
r}(\vec{\Sigma_{1}}+\vec{\Sigma_{2}})\vec{L}+\frac{1}{4m^{2}}\beta_{1}\beta_{2}\alpha_{1i}\alpha_{2j}\nabla_{i}\nabla_{j}V
\label{last}
\end{eqnarray}
The last term in (\ref{last}), though of the order of $v^4$, is a direct product of two odd operators and
therefore does not contribute to the first order perturbation considered in this paper.

\end{document}